\documentclass[twocolumn,secnumarabic,amssymb, nobibnotes, aps, prd]{revtex4-1}

\usepackage{graphicx}

\begin{document}
\title{Testing Lorentz Invariance Using an Odd-Parity Asymmetric Optical Resonator}
\author{Fred N. Baynes}  \email{fred@physics.uwa.edu.au} \author{Andre N. Luiten} \author{Michael E. Tobar}
\affiliation{Frequency Standards and Metrology, The University of Western Australia, 35 Stirling Highway, Crawley, WA 6009, AUSTRALIA}
\date{February 20, 2010}
\begin{abstract}
We present the first experimental test of Lorentz  invariance using the frequency difference between counter-propagating modes in an asymmetric odd-parity optical resonator. This type of test is $\sim10^{4}$ more sensitive to odd-parity and isotropic (scalar) violations of Lorentz invariance than equivalent conventional even-parity experiments due to the asymmetry of the optical resonator. The disadvantages of odd parity resonators have been negated by the use of counter-propagating modes, delivering a high level of immunity  to environmental fluctuations. With a non-rotating experiment our result limits the isotropic Lorentz violating parameter  $\tilde{\kappa}_{tr}$  to  3.4 $\pm$ 6.2 x $10^{-9}$, the best reported constraint from direct measurements. Using this technique the bounds on odd-parity and scalar violations of Lorentz invariance can be improved by many orders of magnitude.
\end{abstract}
\maketitle
\section{Introduction}
The assumption of Lorentz Invariance (LI)  is a vital foundational component of modern physics. This fundamental symmetry of space-time has been rigorously tested since it was first postulated over 100 years ago and all such experiments have so far verified the standard model of particle physics and general relativity to within their precision. Nonetheless, the emergence of quantum gravity and other unified theories\cite{PhysRevD.66.124006,PhysRevD.67.043508,AlanKosteleck?1991545,PhysRevD.39.683}, which hint at possible LI violations, continue to give new impetus to undertake ever more precise tests of LI. 

In order to compare the quality of various experimental tests of LI one can make use of the framework of the minimal Standard Model Extension (SME) by Kostelecky and co-workers \cite{PhysRevD.58.116002}, which parameterizes all possible LI violations by known fields. If an experiment generates a non-zero parameter in this framework then it indicates the degree to which LI is violated. Although the SME is a comprehensive theory with particle, gravity and photon sectors, the experiment reported here is focussed on  the photon sector in which there are 19 independent parameters of the SME. 

In this sector astrophysical observations have determined that the ten parameters representing vacuum birefringence, ($\tilde{\kappa}^{jk}_{e+}$ and $\tilde{\kappa}^{jk}_{o-}$ )  are below  $\sim 10^{-32}$ \cite{PhysRevD.66.056005}. The remaining anisotropic parameters $\tilde{\kappa}^{jk}_{e-}$ and $\tilde{\kappa}^{jk}_{o+}$,  as well as the isotropic parameter $\tilde{\kappa}_{tr}$, have been constrained through laboratory tests using optical or microwave resonators.  The current constraints on the anisotropic even-parity ($\tilde{\kappa}^{jk}_{e-}$ ) parameters are at the level of $10^{-17}$  with the odd-parity coefficients ($\tilde{\kappa}^{jk}_{o+}$  ) at the level of $10^{-13}$ \cite{PhysRevLett.103.090401,PhysRevD.80.105011}. The disparity in these constraints arises because the sensitivity is determined by the symmetry of the sensing apparatus: the most sensitive cavity experiments are based on Michelson-Morley style experiments which are sensitive in leading-order to only the even $\tilde{\kappa}^{jk}_{e-}$ parameters. Hence these parameters are the best constrained of the laboratory-measured SME parameters. The sensitivity of even-parity experiments to odd-parity and isotropic SME coefficients arises solely because of the motion of the earth $v_{\oplus}$ relative to a sun-centered reference frame. For an even-parity experiment the sensitivity to the odd coefficients $\tilde{\kappa}^{jk}_{o+}$ is reduced by the earth's velocity normalized to the speed of light: $\beta = \frac{v_{\oplus}}{c} \simeq 10^{-4}$) \cite{PhysRevD.66.056005}. The sensitivity to the isotropic parameter $\tilde{\kappa}_{tr}$ is reduced further by a factor of $\beta^{2} \simeq 10^{-8}$ \cite{PhysRevD.82.076001}. On the other hand, an odd-parity sensor can be leading order sensitive to the odd SME parameters $\tilde{\kappa}^{jk}_{o}$ \cite{PhysRevD.71.025004,PhysRevD.75.056002}, while only having first order $\beta$ suppression of $\tilde{\kappa}_{tr}$ - see Table \ref{kaps} Hence an asymmetric odd-parity experiment can measure the odd-parity and isotropic SME parameters with enhanced sensitivity compared to previous even-parity Michelson-Morley type experiments.  Here we report results from the first odd-parity optical resonator experiment and we are thus able to provide a constraint on the isotropic parameter $\tilde{\kappa}_{tr}$ with the highest sensitivity yet reported. We further believe that this type of experiment has room for much improvement into the future whereas existing even-parity experiments are near the limit of development and are unlikely to improve by 4 orders of magnitude, limiting the potential for progress in the search for odd-parity and isotropic violations of Lorentz invariance.
\begin{table}[h]
\caption{\label{tab:fonts} Sensitivity to the SME Parameters for different types of resonators  ($\beta \simeq 10^{-4}$)}
\begin{ruledtabular}
\begin{tabular}{lccc}
\label{kaps}
Experiment & \multicolumn{3}{c}{Parameter Sensitivity}  \\
\hline
Even-Parity &   $\sim\tilde{\kappa}^{jk}_{e-}$  & $\sim\beta \tilde{\kappa}^{jk}_{o+}$ & $\sim\beta^{2}  \tilde{\kappa}_{tr}$\\
Odd-Parity  &   $\sim\tilde{\kappa}^{jk}_{o+}$ &  $\sim\beta \tilde{\kappa}_{tr}$ & $\sim\beta^{2} \tilde{\kappa}^{jk}_{e-}$\\
\end{tabular}
\end{ruledtabular}
\end{table}

An analysis of an even-parity rotating microwave resonator experiment designed to test LI \cite{PhysRevD.74.081101} has determined $\tilde{\kappa}_{tr}$ as $15 \pm 7.4 \times 10^{-9}$ \cite{PhysRevD.82.076001}. An alternative means to determine  $\tilde{\kappa}_{tr}$ was obtained using relativistic ion spectroscopy \cite{Reinhardt:2007fk} with sensitivity of $\pm  8.4 \times 10^{-8}$. An odd-parity interferometer has been used to determine  $\tilde{\kappa}_{tr}$ as $-0.03 \pm 3 \times 10^{-7}$, limited by vibrational noise \cite{PhysRevD.80.125024}. A number of other measurements of $\tilde{\kappa}_{tr}$ have been performed based upon astrophysical observations \cite{PhysRevD.78.085026}, collider physics \cite{PhysRevD.80.091901,PhysRevD.80.036010}, or measurements of the electron spin \cite{PhysRevD.74.077901} but these are indirect measurements or contain underlying model assumptions \cite{RevModPhys.83.11}.
\section{Asymmetric Optical Resonator}
\begin{figure}[h]
  \begin{center}  
    \includegraphics[width=0.5 \columnwidth]{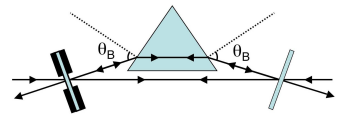}
        \caption{Asymmetric ring resonator with UV fused silica prism at Brewster's angle $\theta_{B}$, showing counter-propagating modes.\label{cav}}
  \end{center}
\end{figure}

The observable in a resonator-based test  of  LI is the normalized frequency shift $\delta \nu/ \nu$ in the resonant frequency of the cavity. As an example, the best  even-parity optical resonator experiment makes use of a rotating block of ULE glass containing two symmetric orthogonal high finesse Fabry-Perot resonators in a heavily isolated and temperature controlled vacuum environment \cite{PhysRevLett.103.090401,PhysRevD.80.105011}. Any violation  of Lorentz invariance will be manifested as modulations in the frequency difference between   the two cavities, related to the rotation of the apparatus. Odd-parity experiments need to break the $180^{\circ}$ rotational symmetry of an even parity experiment. In the case of the experiment described here this asymmetry is achieved by placing a dielectric in one arm of a ring-resonator, see Fig. \ref{cav}. The requirement to include  a dielectric element in the cavity means we cannot simply obtain temperature insensitivity by constructing the cavity from  low thermal expansions materials such as ULE.  We overcome this drawback by sensing the frequency difference between counter-propagating modes, eliminating many of the causes of drift between the cavities. For example, most environmentally-driven changes in the optical path length are common to both counter-propagating modes and thus generate  no effect on the frequency difference between the two modes. This makes the resonator insensitive to environmentally induced fluctuations, which is highlighted by the fact that no temperature control or vibration isolation was required. 
Using the derivation of resonator sensitivity to SME parameters outlined in \cite{PhysRevD.66.056005}, and those which specifically apply to an odd-parity cavity in \cite{PhysRevD.75.056002} and \cite{PhysRevD.71.025004}, the only non-zero term contributing to the observable $\delta \nu/ \nu$ is
\begin{displaymath}
(\mathcal{M}_{DB})^{jk}_{lab}=\text{Re} \left[ -\frac{1}{2\langle U \rangle} \sqrt{ \frac{\epsilon_{0}}{\mu_{0}}} \int_{V} (\textbf{E}^{\ast j}_{0c} \textbf{B}^{k}_{0c}) dV \right]
\end{displaymath}
\noindent
where \textbf{E} and \textbf{B} are the components of the propagating electromagnetic fields and $\langle U \rangle$ is the total energy in the mode. We see thus a secondary benefit of the use of counter-propagating modes; it makes the experiment twice as sensitive to a non-zero SME parameter when compared with an experiment which uses the frequency shifts of a uni-directional beam in an asymmetric cavity. 
\section{The Experiment}
The asymmetric ring cavity was machined out of single aluminium block and is approximately 5 cm $\times$ 5cm, with one of the mirrors was mounted between piezo-electric actuators for cavity length adjustment. The dielectric element is a UV fused silica Brewster angle prism (n=1.44) with a base of 1.7cm.  We are careful to use a Brewster's angle prism to minimize the surface losses together with a low dissipation dielectric material (UV grade fused silica). The measured finesse was 860 with contrast of 0.1 and a free spectral range of 3.85 GHz. Since the resolution of resonator experiments is set by the finesse of the cavity we used relatively high optical power to overcome the modest finesse and ensure optimal conditions for locking to the frequency of the two modes. Fluctuations associated with the high optical power (such as heating and non-linear effects) are once again mitigated by the use of counter propagating modes. We note that the low finesse was not a critical factor in this experiment as the frequency fluctuations in the relevant time domain are dominated  by systematic fluctuations rather than limitations of the frequency locking. 
To monitor changes in the resonant frequency of the two counter-propagating modes the output of a laser is split into two paths which are independently frequency locked to the two fundamental modes counter-propagating in the optical resonator. The experiment has been designed to excite the fundamental mode of the optical resonator and we have experimentally verified that we are locked to the correct modes. This ensures that there is complete spatial overlap of the counter-propagating modes and rejection of optical path-length fluctuations. Higher order modes are not frequency degenerate with the fundamental mode and will not be excited while the laser is frequency locked to the fundamental mode. We use the standard Pound-Drever-Hall (PHD) \cite{black:79} technique to create the error signal required for frequency locking with the required phase modulation being provided by direct modulation of the laser crystal \cite{4993055}.  The use of two acousto-optic modulators (AOM) allows independent frequency locks to each of the counter-propagating modes. The frequency is shifted  by a constant 80 MHz ($\nu_{80MHz}$) in the first path using an AOM in the double pass configuration \cite{donley:063112}. This optical path then passes through a polarization maintaining optical fiber, half-wave plate and polarizer to ensure the correct polarization of light is incident on the cavity. The laser is locked to this resonance using the piezo-electric transducer on the laser. In the second path the locked laser light is sent through a second tunable AOM and this is used to provide the frequency corrections.
\begin{figure}[h]
  \begin{center}  
    \includegraphics[width=1 \columnwidth]{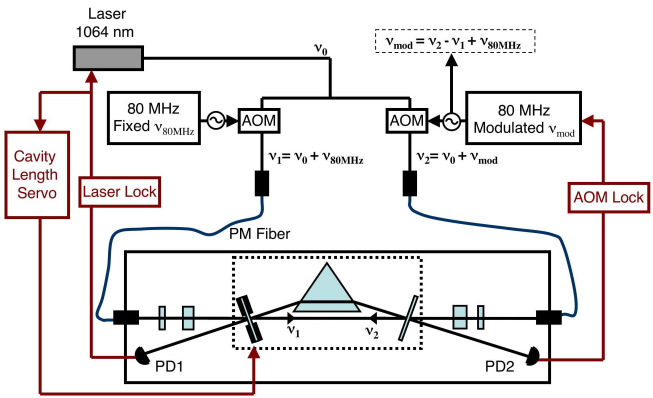}
        \caption{The optical setup, showing the use of two AOMs to eliminate the need for an optical beat note.}
  \label{setup}
  \end{center}
\end{figure}
The second AOM will thus have its frequency locked  at 80MHz plus any frequency difference between the counter-propagating modes ($\nu_{mod}$) - see Fig \ref{setup}. By logging the frequency fluctuations of this   second AOM we can measure the frequency difference between the two counter propagating modes. The AOMs are powered by the amplified output of two  signal generators. The signal generators and counters are all phase-locked to a common 10 MHz signal from a H-Maser. Since any violations of LI will show only in the difference between the resonant frequencies of the two counter propagating modes it is unnecessary to temperature stabilize the cavity to ensure a constant resonant frequency. However, in order to prevent large changes in operating conditions we use the piezo-electric transducers in the cavity and an additional slow loop to keep the laser at a relatively constant frequency.
\section{Data Analysis}
The data was acquired over 50 days from the 19/11/2010 with about 40 days of usable data. Since the experiment is stationary in the laboratory we are searching for frequency modulations synchronous with earth's sidereal phase.  To simplify and increase the speed of the analysis process we  average the data into  20 minute blocks. The raw data is then differentiated with respect to sidereal phase to remove offsets and drifts \cite{PhysRevD.81.022003}. The cavity is orientated in an East-West direction to maximize the sensitivity to possible violations of LI and this leads to sensitivity to only the Cos($\omega_{\oplus} T_{\oplus}  \pm \Omega_{\oplus}T$) terms defined in Table I of \cite{PhysRevD.71.025004}. As the data set only comprises a small section of the year we can apply the short data set approximation \cite{PhysRevLett.95.040404} which assumes a constant annual phase over the duration of the experiment $\Omega_{0}$ and decomposes the signal for LI violations into Sine and Cosine terms with coefficients given in Table \ref{coeff}.
\begin{figure}
  \begin{center}  
    \includegraphics[width=1 \columnwidth]{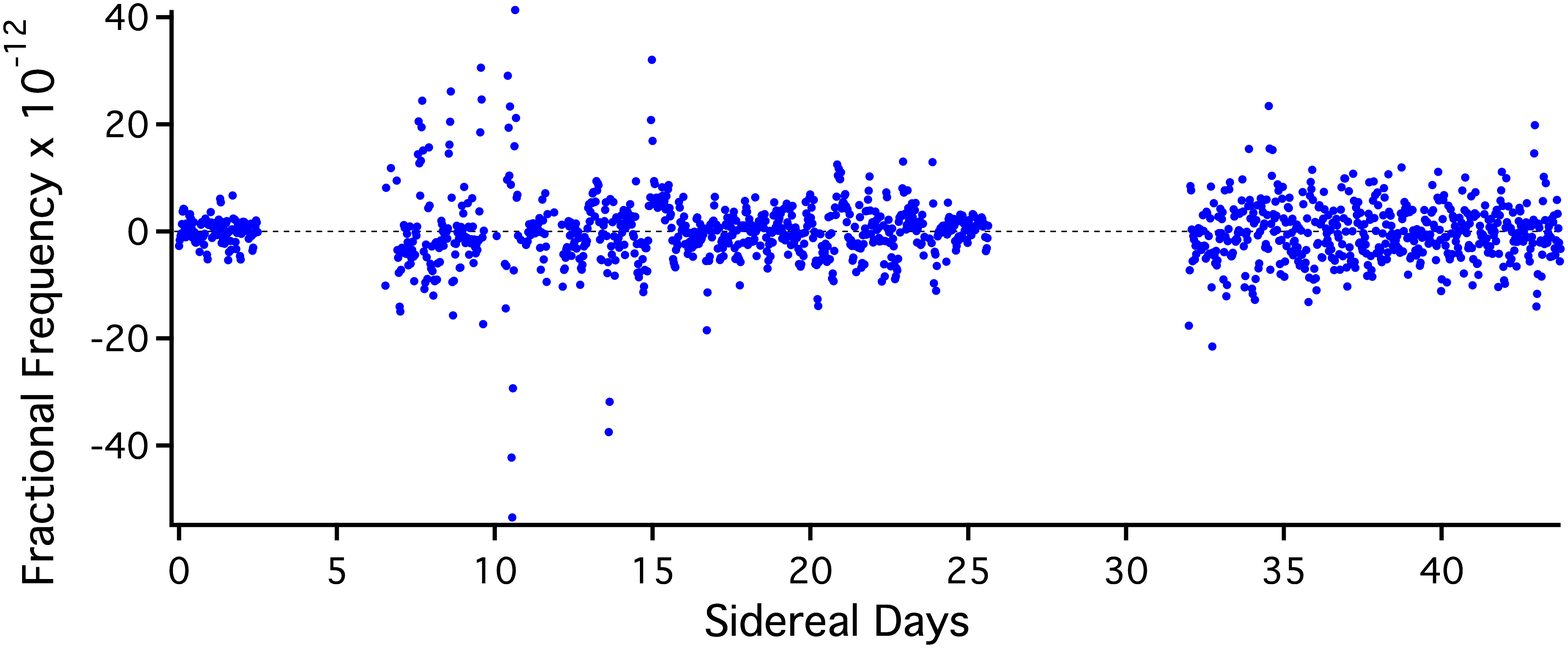}
     \includegraphics[width=1 \columnwidth]{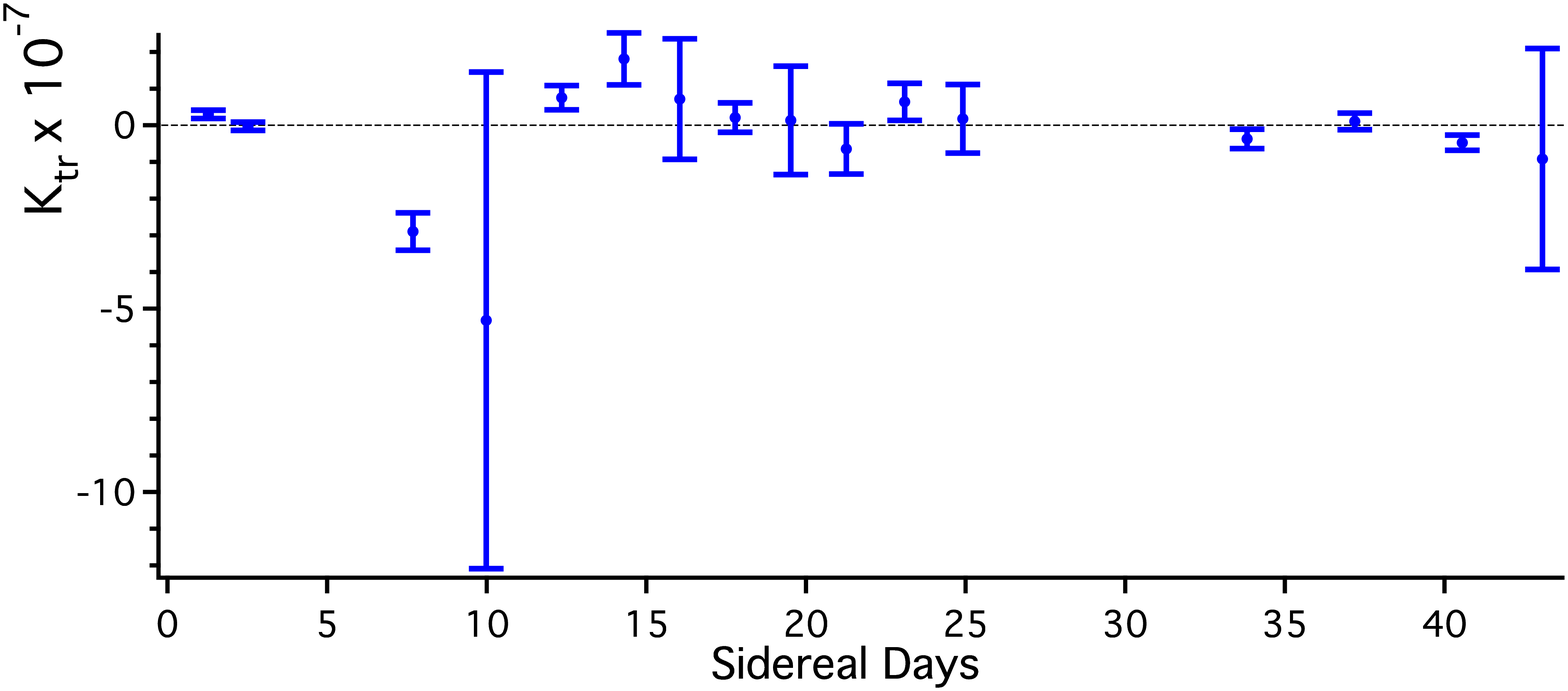}
    \includegraphics[width=1 \columnwidth]{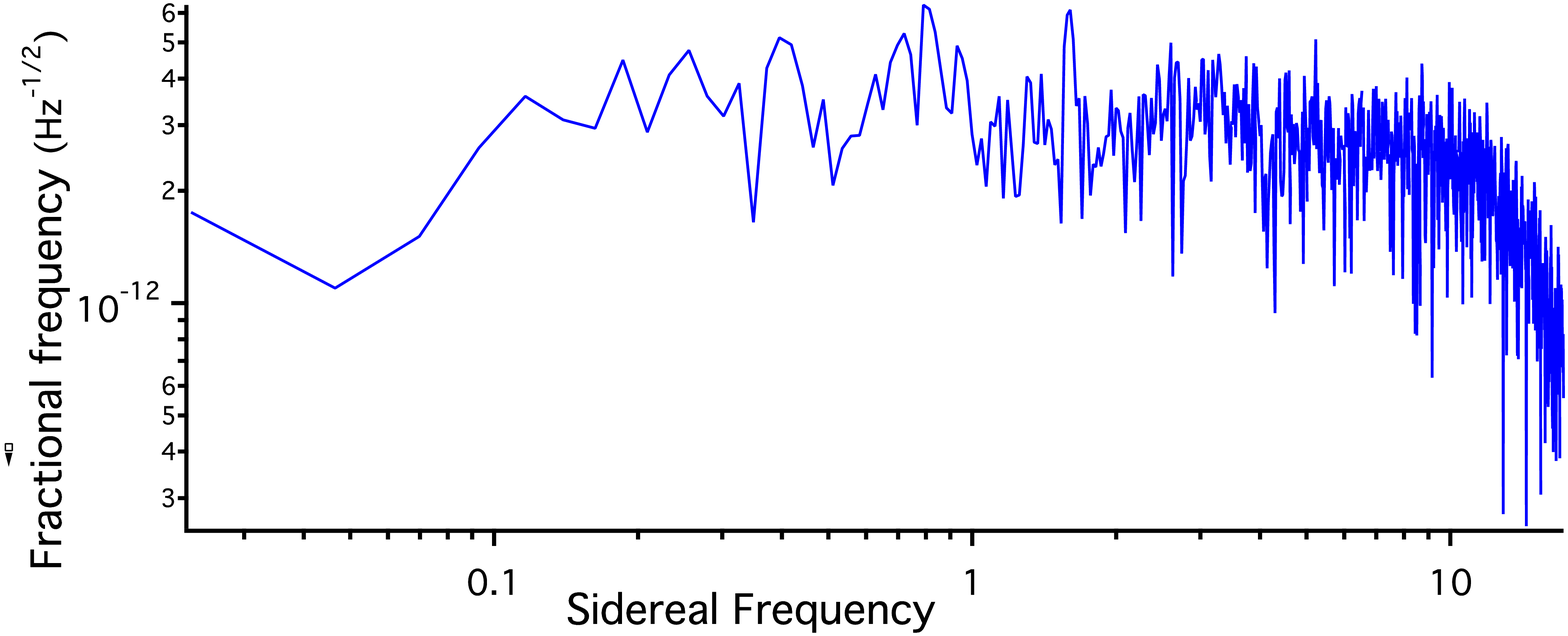}   
        \caption{Difference in resonant frequency of counter propagating modes. Processed time series data (top) and spectral density (bottom). The middle graph are the values obtained for $\tilde{\kappa}_{tr}$ from the data set split up into $\sim$2 day portions   \label{fits}}
  \end{center}
\end{figure}
The co-ordinate system used for analysis in the SME is given in  \cite{PhysRevD.66.056005}, with $T_{\oplus}$ the time since the laboratory frame y axis pointed towards $90^{\circ}$ right ascension, $\eta$ the angle between the celestial orbital plane and the elliptic ($\eta \approx 23.40^{\circ}$) and $\omega_{\oplus}$ is the sidereal frequency. In odd-parity experiments signals for LI violations occur at the rotation frequency (in the case of stationary experiment this corresponds to the sidereal frequency). The data set was divided into sections $\sim$2 days long and the amplitudes of the sin($\omega_{\oplus} \text{T}_{\oplus}) $ and cos($\omega_{\oplus} \text{T}_{\oplus}) $ components are determined using a fitted least squares regression and from these two amplitudes a determination of $\tilde{\kappa}_{tr}$ and $\tilde{\kappa}^{XZ}_{o+}$ are made, and the quoted values are a weighted mean of all the data sets - see Fig. \ref{fits}. The value determined for the odd-parity parameter $\tilde{\kappa}^{XZ}_{o+}$ is $0.7 \pm 1.4 \times 10^{-12}$ which is an order of magnitude above the current limit. For the scalar parameter $\tilde{\kappa}_{tr}$ the result is $3.4 \pm 6.2 \times 10^{-9}$, a new limit on the constraint. The uncompetitive constraint placed on the odd parameter $\tilde{\kappa}^{XZ}_{o+}$ compared to the scalar parameter $\tilde{\kappa}_{tr}$ is because the current best constraints are derived from experiments with different sensitivities (\cite{PhysRevLett.103.090401,PhysRevD.80.105011} and \cite{PhysRevD.82.076001} respectively) and in this first realization of an asymmetric optical resonator we use a non-rotating experiment, giving reduced sensitivity to the odd parameters \cite{PhysRevD.71.025004}.

\begin{table}
\caption{\label{tab:fonts} Sensitivity coefficients of  $\tilde{\kappa}_{tr}$ for this experiment (stationary) using the short data set approximation and differentiated data}
\begin{ruledtabular}
\begin{tabular}{ccc}
Modulation & Coefficient  & Numerical Value\\
\hline
sin($\omega_{\oplus} \text{T}_{\oplus}) $&  $-2  \beta$ \text{cos}($\eta$) cos($\Omega_{0}) \times $& $-2.57\times10^{-5} $cos($\Omega_{0})$\\
&$ \left [(\mathcal{M}_{\text{DB}})^{\text{XZ}}_{\text{lab}}-(\mathcal{M}_{\text{DB}})^{ZX}_{\text{lab}}\right]$&\\
 & &\\
cos($\omega_{\oplus} \text{T}_{\oplus}) $ &   $2 \beta$ sin$(\Omega_{0}) \times $& $2.80\times10^{-5}$sin($\Omega_{0})$\\
&$ \left [(\mathcal{M}_{\text{DB}})^{\text{XZ}}_{\text{lab}}-(\mathcal{M}_{\text{DB}})^{\text{ZX}}_{\text{lab}}\right]$&\\
\label{coeff}
\end{tabular}
\end{ruledtabular}
\end{table}
\section{Discussion of Systematics}
 Although common mode rejection of most environmental effects is a consequence of the counter-propagating mode design, there are nonetheless some systematic effects that afflict this experiment. In the PDH locking scheme unwanted Residual Amplitude Modulation (RAM) co-present with the intended frequency modulation (FM) causes the laser to lock slightly off the centre of resonance \cite{black:79}. In usual PDH systems fluctuations in  RAM will cause frequency fluctuations, although in our approach there is a rejection of this effect if the coupling and finesse of the counter-propagating modes were to be exactly the same.  However,  small alignment and mode-matching differences on the two modes leads to  a small residual sensitivity to the level of RAM. The measured level of alignment fluctuations are consistent with the measured level of frequency fluctuations in this experiment when allowing for the mismatch of contrast on the two modes. Such systematic effects are the major source of instability in the experiment and are a limiting factor in the current constraint on $\tilde{\kappa}_{tr}$. There are non-reciprocal effects associated with the Faraday effect and stress birefringence in the fused silica that will cause a frequency difference between the two counter-propagating modes, related to the presence of magnetic fields in the laboratory \cite{0049-1748-1-5-A05}. Based on measurements of the magnetic field strength and variation near the optical resonator the calculated effect is two orders of magnitude below the uncertainty in $\tilde{\kappa}_{tr}$. The use of counter-propagating modes in a ring resonator means that the device will exhibit a sensitivity to rotational velocity in the plane of the device (Sagnac effect \cite{RevModPhys.39.475,Takahashi:88}). The Sagnac effect would affect the LI result only if the angular velocity or the dimensions of the optical resonator were to fluctuate with a sidereal period but the presence of this systematic effect is more than four orders of magnitude below the uncertainty in $\tilde{\kappa}_{tr}$. These systematic effects are technical limits to the sensitivity of this particular experiment and can be drastically reduced through alignment and temperature control, magnetic shielding and a higher finesse or contrast cavity. 
 
 
\section{Conclusion}
\begin{figure}
  \begin{center}  
    \includegraphics[width=0.8 \columnwidth]{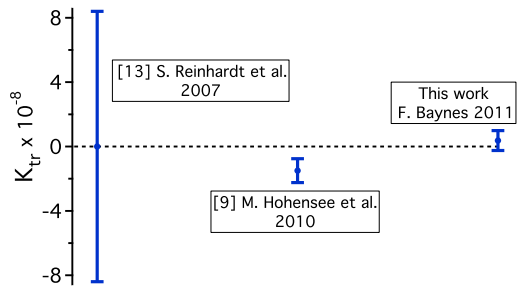}
        \caption{Comparison of  $\tilde{\kappa}_{tr}$ determined by other experiments.}
  \label{comp}
  \end{center}
\end{figure}
An odd parity experiment offers $10^{4}$ times more sensitivity to the odd-parity $\tilde{\kappa}^{jk}_{o+}$ and isotropic  $\tilde{\kappa}_{tr}$ parameters in experimental tests of LI. The value for $\tilde{\kappa}_{tr}$ determined from this experiment is $\tilde{\kappa}_{tr}= 3.4 \pm 6.2 \times 10^{-9}$  $(1\sigma \text{ error})$, the tightest published constraint on $\tilde{\kappa}_{tr}$ to our knowledge. This constraint is more than a factor of 12 better than previous test of LI at optical frequencies \cite{Reinhardt:2007fk} and moderately better than the previous best \cite{PhysRevD.82.076001}. This experiment is the first odd-parity optical resonator experiment and the use of counter propagating modes enables a new constraint on $\tilde{\kappa}_{tr}$ using a non-rotating resonator without temperature control, vibration isolation or vacuum systems. Given the inherent rejection of environmental fluctuations by the counter-propagating modes, sufficiently advanced odd-parity experiments can now approach the sensitivity of the state-of-the-art Michelson-Morely type even parity experiments. This would enable bounds on odd-parity $\tilde{\kappa}^{jk}_{o+}$ and isotropic $\tilde{\kappa}_{tr}$ to increase by up to 4 orders of magnitude, making  odd-parity optical resonators an important experimental tool in the continuing search for violations of Lorentz Invariance. This work was supported by the Australian Research Council.



\begin{thebibliography}{10}%
\makeatletter
\providecommand \@ifxundefined [1]{%
 \ifx #1\undefined \expandafter \@firstoftwo
 \else \expandafter \@secondoftwo
\fi
}%
\providecommand \@ifnum [1]{%
 \ifnum #1\expandafter \@firstoftwo
 \else \expandafter \@secondoftwo
\fi
}%
\providecommand \enquote [1]{``#1''}%
\providecommand \bibnamefont  [1]{#1}%
\providecommand \bibfnamefont [1]{#1}%
\providecommand \citenamefont [1]{#1}%
\providecommand\href[0]{\@sanitize\@href}%
\providecommand\@href[1]{\endgroup\@@startlink{#1}\endgroup\@@href}%
\providecommand\@@href[1]{#1\@@endlink}%
\providecommand \@sanitize [0]{\begingroup\catcode`\&12\catcode`\#12\relax}%
\@ifxundefined \pdfoutput {\@firstoftwo}{%
 \@ifnum{\z@=\pdfoutput}{\@firstoftwo}{\@secondoftwo}%
}{%
 \providecommand\@@startlink[1]{\leavevmode}%
 \providecommand\@@endlink[0]{}%
}{%
 \providecommand\@@startlink[1]{%
  \leavevmode
  \pdfstartlink
   attr{/Border[0 0 1 ]/H/I/C[0 1 1]}%
   user{/Subtype/Link/A<</Type/Action/S/URI/URI(#1)>>}%
  \relax
 }%
 \providecommand\@@endlink[0]{\pdfendlink}%
}%
\providecommand \url  [0]{\begingroup\@sanitize \@url }%
\providecommand \@url [1]{\endgroup\@href {#1}{\urlprefix}}%
\providecommand \urlprefix [0]{URL }%
\providecommand \Eprint[0]{\href }%
\@ifxundefined \urlstyle {%
  \providecommand \doi [1]{doi:\discretionary{}{}{}#1}%
}{%
  \providecommand \doi [0]{doi:\discretionary{}{}{}\begingroup
  \urlstyle{rm}\Url }%
}%
\providecommand \doibase [0]{http://dx.doi.org/}%
\providecommand \Doi[1]{\href{\doibase#1}}%
\providecommand \selectlanguage [0]{\@gobble}%
\providecommand \bibinfo [0]{\@secondoftwo}%
\providecommand \bibfield [0]{\@secondoftwo}%
\providecommand \translation [1]{[#1]}%
\providecommand \BibitemOpen[0]{}%
\providecommand \bibitemStop [0]{}%
\providecommand \bibitemNoStop [0]{.\EOS\space}%
\providecommand \EOS [0]{\spacefactor3000\relax}%
\providecommand \BibitemShut [1]{\csname bibitem#1\endcsname}%
\bibitem{PhysRevD.66.124006}%
  \BibitemOpen
  \bibfield{author}{%
  \bibinfo {author} {\bibfnamefont{J.}~\bibnamefont{Alfaro}}, \bibinfo {author}
  {\bibfnamefont{H.~A.}\ \bibnamefont{Morales-T\'ecotl}},\ and\ \bibinfo
  {author} {\bibfnamefont{L.~F.}\ \bibnamefont{Urrutia}},\ }%
  \bibfield{journal}{%
  \bibinfo {journal} {Phys. Rev. D}\ }%
  \textbf{\bibinfo {volume} {66}},\ \bibinfo {pages} {124006} (\bibinfo {month}
  {Dec}\ \bibinfo {year} {2002})\BibitemShut{NoStop}%
\bibitem{PhysRevD.67.043508}%
  \BibitemOpen
  \bibfield{author}{%
  \bibinfo {author} {\bibfnamefont{J.~D.}\ \bibnamefont{Bjorken}},\ }%
  \bibfield{journal}{%
  \bibinfo {journal} {Phys. Rev. D}\ }%
  \textbf{\bibinfo {volume} {67}},\ \bibinfo {pages} {043508} (\bibinfo {month}
  {Feb}\ \bibinfo {year} {2003})\BibitemShut{NoStop}%
\bibitem{AlanKosteleck?1991545}%
  \BibitemOpen
  \bibfield{author}{%
  \bibinfo {author} {\bibfnamefont{V.~A.}\ \bibnamefont{Kosteleck\'y}}\ and\
  \bibinfo {author} {\bibfnamefont{R.}~\bibnamefont{Potting}},\ }%
  \bibfield{journal}{%
  \bibinfo {journal} {Nuclear Physics B}\ }%
  \textbf{\bibinfo {volume} {359}},\ \bibinfo {pages} {545 } (\bibinfo {year}
  {1991}),\ ISSN \bibinfo {issn} {0550-3213}\BibitemShut{NoStop}%
\bibitem{PhysRevD.39.683}%
  \BibitemOpen
  \bibfield{author}{%
  \bibinfo {author} {\bibfnamefont{V.~A.}\ \bibnamefont{Kosteleck\'y}}\ and\
  \bibinfo {author} {\bibfnamefont{S.}~\bibnamefont{Samuel}},\ }%
  \bibfield{journal}{%
  \bibinfo {journal} {Phys. Rev. D}\ }%
  \textbf{\bibinfo {volume} {39}},\ \bibinfo {pages} {683} (\bibinfo {month}
  {Jan}\ \bibinfo {year} {1989})\BibitemShut{NoStop}%
\bibitem{PhysRevD.58.116002}%
  \BibitemOpen
  \bibfield{author}{%
  \bibinfo {author} {\bibfnamefont{D.}~\bibnamefont{Colladay}}\ and\ \bibinfo
  {author} {\bibfnamefont{V.~A.}\ \bibnamefont{Kosteleck\'y}},\ }%
  \bibfield{journal}{%
  \bibinfo {journal} {Phys. Rev. D}\ }%
  \textbf{\bibinfo {volume} {58}},\ \bibinfo {pages} {116002} (\bibinfo {month}
  {Oct}\ \bibinfo {year} {1998})\BibitemShut{NoStop}%
\bibitem{PhysRevD.66.056005}%
  \BibitemOpen
  \bibfield{author}{%
  \bibinfo {author} {\bibfnamefont{V.~A.}\ \bibnamefont{Kosteleck\'y}}\ and\
  \bibinfo {author} {\bibfnamefont{M.}~\bibnamefont{Mewes}},\ }%
  \bibfield{journal}{%
  \bibinfo {journal} {Phys. Rev. D}\ }%
  \textbf{\bibinfo {volume} {66}},\ \bibinfo {pages} {056005} (\bibinfo {month}
  {Sep}\ \bibinfo {year} {2002})\BibitemShut{NoStop}%
\bibitem{PhysRevLett.103.090401}%
  \BibitemOpen
  \bibfield{author}{%
  \bibinfo {author} {\bibfnamefont{C.}~\bibnamefont{Eisele}}, \bibinfo {author}
  {\bibfnamefont{A.~Y.}\ \bibnamefont{Nevsky}},\ and\ \bibinfo {author}
  {\bibfnamefont{S.}~\bibnamefont{Schiller}},\ }%
  \bibfield{journal}{%
  \bibinfo {journal} {Phys. Rev. Lett.}\ }%
  \textbf{\bibinfo {volume} {103}},\ \bibinfo {pages} {090401} (\bibinfo
  {month} {Aug}\ \bibinfo {year} {2009})\BibitemShut{NoStop}%
\bibitem{PhysRevD.80.105011}%
  \BibitemOpen
  \bibfield{author}{%
  \bibinfo {author} {\bibfnamefont{S.}~\bibnamefont{Herrmann}}, \bibinfo
  {author} {\bibfnamefont{A.}~\bibnamefont{Senger}}, \bibinfo {author}
  {\bibfnamefont{K.}~\bibnamefont{M\"ohle}}, \bibinfo {author}
  {\bibfnamefont{M.}~\bibnamefont{Nagel}}, \bibinfo {author}
  {\bibfnamefont{E.~V.}\ \bibnamefont{Kovalchuk}},\ and\ \bibinfo {author}
  {\bibfnamefont{A.}~\bibnamefont{Peters}},\ }%
  \bibfield{journal}{%
  \bibinfo {journal} {Phys. Rev. D}\ }%
  \textbf{\bibinfo {volume} {80}},\ \bibinfo {pages} {105011} (\bibinfo {month}
  {Nov}\ \bibinfo {year} {2009})\BibitemShut{NoStop}%
\bibitem{PhysRevD.82.076001}%
  \BibitemOpen
  \bibfield{author}{%
  \bibinfo {author} {\bibfnamefont{M.~A.}\ \bibnamefont{Hohensee}}, \bibinfo
  {author} {\bibfnamefont{P.~L.}\ \bibnamefont{Stanwix}}, \bibinfo {author}
  {\bibfnamefont{M.~E.}\ \bibnamefont{Tobar}}, \bibinfo {author}
  {\bibfnamefont{S.~R.}\ \bibnamefont{Parker}}, \bibinfo {author}
  {\bibfnamefont{D.~F.}\ \bibnamefont{Phillips}},\ and\ \bibinfo {author}
  {\bibfnamefont{R.~L.}\ \bibnamefont{Walsworth}},\ }%
  \bibfield{journal}{%
  \bibinfo {journal} {Phys. Rev. D}\ }%
  \textbf{\bibinfo {volume} {82}},\ \bibinfo {pages} {076001} (\bibinfo {month}
  {Oct}\ \bibinfo {year} {2010})\BibitemShut{NoStop}%
\bibitem{PhysRevD.71.025004}%
  \BibitemOpen
  \bibfield{author}{%
  \bibinfo {author} {\bibfnamefont{M.~E.}\ \bibnamefont{Tobar}}, \bibinfo
  {author} {\bibfnamefont{P.}~\bibnamefont{Wolf}}, \bibinfo {author}
  {\bibfnamefont{A.}~\bibnamefont{Fowler}},\ and\ \bibinfo {author}
  {\bibfnamefont{J.~G.}\ \bibnamefont{Hartnett}},\ }%
  \bibfield{journal}{%
  \bibinfo {journal} {Phys. Rev. D}\ }%
  \textbf{\bibinfo {volume} {71}},\ \bibinfo {pages} {025004} (\bibinfo {month}
  {Jan}\ \bibinfo {year} {2005})\BibitemShut{NoStop}%
\bibitem{PhysRevD.75.056002}%
  \BibitemOpen
  \bibfield{author}{%
  \bibinfo {author} {\bibfnamefont{M.}~\bibnamefont{Mewes}}\ and\ \bibinfo
  {author} {\bibfnamefont{A.}~\bibnamefont{Petroff}},\ }%
  \bibfield{journal}{%
  \bibinfo {journal} {Phys. Rev. D}\ }%
  \textbf{\bibinfo {volume} {75}},\ \bibinfo {pages} {056002} (\bibinfo {month}
  {Mar}\ \bibinfo {year} {2007})\BibitemShut{NoStop}%
\bibitem{PhysRevD.74.081101}%
  \BibitemOpen
  \bibfield{author}{%
  \bibinfo {author} {\bibfnamefont{P.~L.}\ \bibnamefont{Stanwix}}, \bibinfo
  {author} {\bibfnamefont{M.~E.}\ \bibnamefont{Tobar}}, \bibinfo {author}
  {\bibfnamefont{P.}~\bibnamefont{Wolf}}, \bibinfo {author}
  {\bibfnamefont{C.~R.}\ \bibnamefont{Locke}},\ and\ \bibinfo {author}
  {\bibfnamefont{E.~N.}\ \bibnamefont{Ivanov}},\ }%
  \bibfield{journal}{%
  \bibinfo {journal} {Phys. Rev. D}\ }%
  \textbf{\bibinfo {volume} {74}},\ \bibinfo {pages} {081101} (\bibinfo {month}
  {Oct}\ \bibinfo {year} {2006})\BibitemShut{NoStop}%
\bibitem{Reinhardt:2007fk}%
  \BibitemOpen
  \bibfield{author}{%
  \bibinfo {author} {\bibfnamefont{S.}~\bibnamefont{Reinhardt}}, \bibinfo
  {author} {\bibfnamefont{G.}~\bibnamefont{Saathoff}}, \bibinfo {author}
  {\bibfnamefont{H.}~\bibnamefont{Buhr}}, \bibinfo {author}
  {\bibfnamefont{L.~A.}\ \bibnamefont{Carlson}}, \bibinfo {author}
  {\bibfnamefont{A.}~\bibnamefont{Wolf}}, \bibinfo {author}
  {\bibfnamefont{D.}~\bibnamefont{Schwalm}}, \bibinfo {author}
  {\bibfnamefont{S.}~\bibnamefont{Karpuk}}, \bibinfo {author}
  {\bibfnamefont{C.}~\bibnamefont{Novotny}}, \bibinfo {author}
  {\bibfnamefont{G.}~\bibnamefont{Huber}}, \bibinfo {author}
  {\bibfnamefont{M.}~\bibnamefont{Zimmermann}}, \bibinfo {author}
  {\bibfnamefont{R.}~\bibnamefont{Holzwarth}}, \bibinfo {author}
  {\bibfnamefont{T.}~\bibnamefont{Udem}}, \bibinfo {author}
  {\bibfnamefont{T.~W.}\ \bibnamefont{Hansch}},\ and\ \bibinfo {author}
  {\bibfnamefont{G.}~\bibnamefont{Gwinner}},\ }%
  \bibfield{journal}{%
  \bibinfo {journal} {Nat Phys}\ }%
  \textbf{\bibinfo {volume} {3}},\ \bibinfo {pages} {861} (\bibinfo {month}
  {12}\ \bibinfo {year} {2007})\BibitemShut{NoStop}%
\bibitem{PhysRevD.80.125024}%
  \BibitemOpen
  \bibfield{author}{%
  \bibinfo {author} {\bibfnamefont{M.~E.}\ \bibnamefont{Tobar}}, \bibinfo
  {author} {\bibfnamefont{E.~N.}\ \bibnamefont{Ivanov}}, \bibinfo {author}
  {\bibfnamefont{P.~L.}\ \bibnamefont{Stanwix}}, \bibinfo {author}
  {\bibfnamefont{J.-M.~G.}\ \bibnamefont{le~Floch}},\ and\ \bibinfo {author}
  {\bibfnamefont{J.~G.}\ \bibnamefont{Hartnett}},\ }%
  \bibfield{journal}{%
  \bibinfo {journal} {Phys. Rev. D}\ }%
  \textbf{\bibinfo {volume} {80}},\ \bibinfo {pages} {125024} (\bibinfo {month}
  {Dec}\ \bibinfo {year} {2009})\BibitemShut{NoStop}%
\bibitem{PhysRevD.78.085026}%
  \BibitemOpen
  \bibfield{author}{%
  \bibinfo {author} {\bibfnamefont{F.~R.}\ \bibnamefont{Klinkhamer}}\ and\
  \bibinfo {author} {\bibfnamefont{M.}~\bibnamefont{Schreck}},\ }%
  \bibfield{journal}{%
  \bibinfo {journal} {Phys. Rev. D}\ }%
  \textbf{\bibinfo {volume} {78}},\ \bibinfo {pages} {085026} (\bibinfo {month}
  {Oct}\ \bibinfo {year} {2008})\BibitemShut{NoStop}%
\bibitem{PhysRevD.80.091901}%
  \BibitemOpen
  \bibfield{author}{%
  \bibinfo {author} {\bibfnamefont{B.}~\bibnamefont{Altschul}},\ }%
  \bibfield{journal}{%
  \bibinfo {journal} {Phys. Rev. D}\ }%
  \textbf{\bibinfo {volume} {80}},\ \bibinfo {pages} {091901} (\bibinfo {month}
  {Nov}\ \bibinfo {year} {2009})\BibitemShut{NoStop}%
\bibitem{PhysRevD.80.036010}%
  \BibitemOpen
  \bibfield{author}{%
  \bibinfo {author} {\bibfnamefont{M.~A.}\ \bibnamefont{Hohensee}}, \bibinfo
  {author} {\bibfnamefont{R.}~\bibnamefont{Lehnert}}, \bibinfo {author}
  {\bibfnamefont{D.~F.}\ \bibnamefont{Phillips}},\ and\ \bibinfo {author}
  {\bibfnamefont{R.~L.}\ \bibnamefont{Walsworth}},\ }%
  \bibfield{journal}{%
  \bibinfo {journal} {Phys. Rev. D}\ }%
  \textbf{\bibinfo {volume} {80}},\ \bibinfo {pages} {036010} (\bibinfo {month}
  {Aug}\ \bibinfo {year} {2009})\BibitemShut{NoStop}%
\bibitem{PhysRevD.74.077901}%
  \BibitemOpen
  \bibfield{author}{%
  \bibinfo {author} {\bibfnamefont{C.~D.}\ \bibnamefont{Carone}}, \bibinfo
  {author} {\bibfnamefont{M.}~\bibnamefont{Sher}},\ and\ \bibinfo {author}
  {\bibfnamefont{M.}~\bibnamefont{Vanderhaeghen}},\ }%
  \bibfield{journal}{%
  \bibinfo {journal} {Phys. Rev. D}\ }%
  \textbf{\bibinfo {volume} {74}},\ \bibinfo {pages} {077901} (\bibinfo {month}
  {Oct}\ \bibinfo {year} {2006})\BibitemShut{NoStop}%
\bibitem{RevModPhys.83.11}%
  \BibitemOpen
  \bibfield{author}{%
  \bibinfo {author} {\bibfnamefont{V.~A.}\ \bibnamefont{Kosteleck\'y}}\ and\
  \bibinfo {author} {\bibfnamefont{N.}~\bibnamefont{Russell}},\ }%
  \bibfield{journal}{%
  \bibinfo {journal} {Rev. Mod. Phys.}\ }%
  \textbf{\bibinfo {volume} {83}},\ \bibinfo {pages} {11} (\bibinfo {month}
  {Mar}\ \bibinfo {year} {2011})\BibitemShut{NoStop}%
\bibitem{black:79}%
  \BibitemOpen
  \bibfield{author}{%
  \bibinfo {author} {\bibfnamefont{E.~D.}\ \bibnamefont{Black}},\ }%
  \bibfield{journal}{%
  \bibinfo {journal} {American Journal of Physics}\ }%
  \textbf{\bibinfo {volume} {69}},\ \bibinfo {pages} {79} (\bibinfo {year}
  {2001})\BibitemShut{NoStop}%
\bibitem{4993055}%
  \BibitemOpen
  \bibfield{author}{%
  \bibinfo {author} {\bibfnamefont{G.}~\bibnamefont{Cantatore}}, \bibinfo
  {author} {\bibfnamefont{F.~D.}\ \bibnamefont{Valle}}, \bibinfo {author}
  {\bibfnamefont{E.}~\bibnamefont{Milotti}}, \bibinfo {author}
  {\bibfnamefont{P.}~\bibnamefont{Pace}}, \bibinfo {author}
  {\bibfnamefont{E.}~\bibnamefont{Zavattini}}, \bibinfo {author}
  {\bibfnamefont{E.}~\bibnamefont{Polacco}}, \bibinfo {author}
  {\bibfnamefont{F.}~\bibnamefont{Perrone}}, \bibinfo {author}
  {\bibfnamefont{C.}~\bibnamefont{Rizzo}}, \bibinfo {author}
  {\bibfnamefont{G.}~\bibnamefont{Zavattini}},\ and\ \bibinfo {author}
  {\bibfnamefont{G.}~\bibnamefont{Ruoso}},\ }%
  \bibfield{journal}{%
  \bibinfo {journal} {Rev. Sci. Inst.}\ }%
  \textbf{\bibinfo {volume} {66}},\ \bibinfo {pages} {2785 } (\bibinfo {month}
  {Apr.}\ \bibinfo {year} {1995}),\ ISSN \bibinfo {issn}
  {0034-6748}\BibitemShut{NoStop}%
\bibitem{donley:063112}%
  \BibitemOpen
  \bibfield{author}{%
  \bibinfo {author} {\bibfnamefont{E.~A.}\ \bibnamefont{Donley}}, \bibinfo
  {author} {\bibfnamefont{T.~P.}\ \bibnamefont{Heavner}}, \bibinfo {author}
  {\bibfnamefont{F.}~\bibnamefont{Levi}}, \bibinfo {author}
  {\bibfnamefont{M.~O.}\ \bibnamefont{Tataw}},\ and\ \bibinfo {author}
  {\bibfnamefont{S.~R.}\ \bibnamefont{Jefferts}},\ }%
  \bibfield{journal}{%
  \bibinfo {journal} {Rev. Sci. Inst.}\ }%
  \textbf{\bibinfo {volume} {76}},\ \bibinfo {eid} {063112} (\bibinfo {year}
  {2005})\BibitemShut{NoStop}%
\bibitem{PhysRevD.81.022003}%
  \BibitemOpen
  \bibfield{author}{%
  \bibinfo {author} {\bibfnamefont{M.~E.}\ \bibnamefont{Tobar}}, \bibinfo
  {author} {\bibfnamefont{P.}~\bibnamefont{Wolf}}, \bibinfo {author}
  {\bibfnamefont{S.}~\bibnamefont{Bize}}, \bibinfo {author}
  {\bibfnamefont{G.}~\bibnamefont{Santarelli}},\ and\ \bibinfo {author}
  {\bibfnamefont{V.}~\bibnamefont{Flambaum}},\ }%
  \bibfield{journal}{%
  \bibinfo {journal} {Phys. Rev. D}\ }%
  \textbf{\bibinfo {volume} {81}},\ \bibinfo {pages} {022003} (\bibinfo {month}
  {Jan}\ \bibinfo {year} {2010})\BibitemShut{NoStop}%
\bibitem{PhysRevLett.95.040404}%
  \BibitemOpen
  \bibfield{author}{%
  \bibinfo {author} {\bibfnamefont{P.~L.}\ \bibnamefont{Stanwix}}, \bibinfo
  {author} {\bibfnamefont{M.~E.}\ \bibnamefont{Tobar}}, \bibinfo {author}
  {\bibfnamefont{P.}~\bibnamefont{Wolf}}, \bibinfo {author}
  {\bibfnamefont{M.}~\bibnamefont{Susli}}, \bibinfo {author}
  {\bibfnamefont{C.~R.}\ \bibnamefont{Locke}}, \bibinfo {author}
  {\bibfnamefont{E.~N.}\ \bibnamefont{Ivanov}}, \bibinfo {author}
  {\bibfnamefont{J.}~\bibnamefont{Winterflood}},\ and\ \bibinfo {author}
  {\bibfnamefont{F.}~\bibnamefont{van Kann}},\ }%
  \bibfield{journal}{%
  \bibinfo {journal} {Phys. Rev. Lett.}\ }%
  \textbf{\bibinfo {volume} {95}},\ \bibinfo {pages} {040404} (\bibinfo {month}
  {Jul}\ \bibinfo {year} {2005})\BibitemShut{NoStop}%
\bibitem{0049-1748-1-5-A05}%
  \BibitemOpen
  \bibfield{author}{%
  \bibinfo {author} {\bibfnamefont{V.~I.}\ \bibnamefont{Chernen'kii}},\ }%
  \bibfield{journal}{%
  \bibinfo {journal} {Soviet Journal of Quantum Electronics}\ }%
  \textbf{\bibinfo {volume} {1}},\ \bibinfo {pages} {472} (\bibinfo {year}
  {1972})\BibitemShut{NoStop}%
\bibitem{RevModPhys.39.475}%
  \BibitemOpen
  \bibfield{author}{%
  \bibinfo {author} {\bibfnamefont{E.~J.}\ \bibnamefont{Post}},\ }%
  \bibfield{journal}{%
  \bibinfo {journal} {Rev. Mod. Phys.}\ }%
  \textbf{\bibinfo {volume} {39}},\ \bibinfo {pages} {475} (\bibinfo {month}
  {Apr}\ \bibinfo {year} {1967})\BibitemShut{NoStop}%
\bibitem{Takahashi:88}%
  \BibitemOpen
  \bibfield{author}{%
  \bibinfo {author} {\bibfnamefont{M.}~\bibnamefont{Takahashi}}, \bibinfo
  {author} {\bibfnamefont{S.}~\bibnamefont{Tai}}, \bibinfo {author}
  {\bibfnamefont{K.}~\bibnamefont{Kyuma}},\ and\ \bibinfo {author}
  {\bibfnamefont{K.}~\bibnamefont{Hamanaka}},\ }%
  \bibfield{journal}{%
  \bibinfo {journal} {Opt. Lett.}\ }%
  \textbf{\bibinfo {volume} {13}},\ \bibinfo {pages} {236} (\bibinfo {month}
  {Mar}\ \bibinfo {year} {1988})\BibitemShut{NoStop}%
\end{thebibliography}

%

\end{document}